\documentclass[amssymb,prb,twocolumn,superscriptaddress,floats,showkeys,showpacs]{revtex4-1}

\usepackage{float}
\usepackage[caption=false]{subfig}
\usepackage{ulem}
\usepackage[euler]{textgreek}
\usepackage{mathrsfs}
\usepackage[T1]{fontenc}
\usepackage{bm}
\usepackage{graphicx}
\usepackage{natbib}
\usepackage{amsmath}
\usepackage{textcomp}
\usepackage{epstopdf}
\usepackage{hyperref} 
\hypersetup{colorlinks,citecolor=blue, filecolor=blue ,linkcolor=blue , urlcolor=blue, pdftex}
\usepackage{sidecap}

\begin{document}

\title{Tuning carrier concentration in a superacid treated MoS$_2$ monolayer}

\author{M. R. Molas}
\email{maciej.molas@fuw.edu.pl}
\affiliation{Institute of Experimental Physics, Faculty of Physics, University of Warsaw, ul. Pasteura 5, 02-093 Warszawa, Poland}
\affiliation{Laboratoire National des Champs Magn\'etiques Intenses, CNRS-UGA-UPS-INSA-EMFL, 25, avenue des Martyrs, 38042 Grenoble, France}
\author{K. Go\l{}asa}
\affiliation{Institute of Experimental Physics, Faculty of Physics, University of Warsaw, ul. Pasteura 5, 02-093 Warszawa, Poland}
\author{\L{}. Bala}
\affiliation{Institute of Experimental Physics, Faculty of Physics, University of Warsaw, ul. Pasteura 5, 02-093 Warszawa, Poland}
\affiliation{Laboratoire National des Champs Magn\'etiques Intenses, CNRS-UGA-UPS-INSA-EMFL, 25, avenue des Martyrs, 38042 Grenoble, France}
\author{K. Nogajewski}
\affiliation{Institute of Experimental Physics, Faculty of Physics, University of Warsaw, ul. Pasteura 5, 02-093 Warszawa, Poland}
\affiliation{Laboratoire National des Champs Magn\'etiques Intenses, CNRS-UGA-UPS-INSA-EMFL, 25, avenue des Martyrs, 38042 Grenoble, France}
\author{M. Bartos}
\affiliation{Laboratoire National des Champs Magn\'etiques Intenses, CNRS-UGA-UPS-INSA-EMFL, 25, avenue des Martyrs, 38042 Grenoble, France}
\author{M. Potemski}
\affiliation{Institute of Experimental Physics, Faculty of Physics, University of Warsaw, ul. Pasteura 5, 02-093 Warszawa, Poland}
\affiliation{Laboratoire National des Champs Magn\'etiques Intenses, CNRS-UGA-UPS-INSA-EMFL, 25, avenue des Martyrs, 38042 Grenoble, France}
\author{A. Babi\'nski}
\email{adam.babinski@fuw.edu.pl}
\affiliation{Institute of Experimental Physics, Faculty of Physics, University of Warsaw, ul. Pasteura 5, 02-093 Warszawa, Poland}

\begin{abstract}
The effect of bis(trifluoromethane) sulfonimide (TFSI, superacid) treatment on the optical properties of MoS$_2$ monolayers is investigated by means of
photoluminescence, reflectance contrast and Raman scattering spectroscopy employed in a 
broad temperature range. It is shown that when applied multiple times, the treatment results in progressive 
quenching of the trion emission/absorption and in the redshift of the neutral exciton emission/absorption 
associated with both the A and B excitonic resonances. Based on this evolution, a trion complex related to the B exciton in monolayer MoS$_2$ is unambiguously identified. A defect-related emission observed 
at low temperatures also disappears from the spectrum as a result of the treatment.
Our observations are attributed to effective passivation of defects on the MoS$_{2}$ monolayer surface.
The passivation reduces the carrier density, which in turn affects the out-of-plane 
electric field in the sample. The observed tuning of the carrier concentration strongly influences also 
the Raman scattering in the MoS$_2$ monolayer. An enhancement of Raman scattering at resonant excitation in the vicinity of the A neutral exciton is clearly seen for both the out-of-plane 
A$_1^{'}$ and in-plane E$^{'}$ modes. On the contrary, when the excitation is in 
resonance with a corresponding trion, the Raman scattering features become hardly visible. These results 
confirm the role of the excitonic charge state plays in the resonance effect of the excitation energy on 
the Raman scattering in transition metal dichalcogenides.
	
\end{abstract}


\maketitle

\section{Introduction \label{sec:Intro}}

	Molybdenum disulfide (MoS$_2$) is the best known representative of semiconducting transition metal dichalcogenides (\mbox{S-TMDs}),
which have recently attracted considerable attention due to their unique electronic
structures and corresponding optical properties~\cite{novoselov2005,wang2012,koperski,Wang2018}.
Like other members of the S-TMD family, MoS$_2$ transforms from indirect- to direct-band-gap 
semiconductor when thinned down from the bulk form to the monolayer (ML) limit~\cite{mak2010,splendiani2010}. 
As efficient light emitters, S-TMD monolayers are considered to be very promising building blocks of 
novel optoelectronic devices~\cite{Yin2012,Baugher2014,Ross2014,Withers2015,binder2017}.

The potential of S-TMDs is related to the their specific crystal structure which comprises strongly
bound metal and chalcogen atoms arranged into one-molecule-thick single layers, which in $N$-layer crystals are stacked one on top of another and kept together by weak van der Waals interaction. In the ideal case of defect-free material, one of the features of such a structure is the absence of dangling 
bonds at the terminal chalcogen layers of each individual S-TMD monolayer. In practice, in order to profit from the unique properties of thin S-TMD films,
it is absolutely crucial to take proper care of the quality and cleanliness of their surface. This
means, in particular, keeping the surface free from chemical residues and avoiding defects which can 
negatively affect the desired optical properties of S-TMD layers. Recently, it has been reported that encapsulation 
of S-TMD monolayers in hexagonal boron nitride leads to suppression of the inhomogeneous 
contribution to the linewidths of excitonic resonances~\cite{cadizMoS2,wierzbowski2017,wang2017,vaclavkova}. Another approach to heal the surface of $N$-layer S-TMDs is to subject them to a specific chemical treatment. Notably, Amani $et$~$al.$~\cite{amaniScience} have shown that treating MoS$_2$ MLs with bis(trifluoromethane) sulfonimide (TFSI),
referred to in what follows as a superacid, results in a considerable increase of the related photoluminescence (PL) intensity.
The reported PL intensity growth amounts to about two orders of magnitude at room temperature. 
Subsequent works~\cite{AmaniNano2016,AmaniACS2016,Kim2017,cadizacid,kiriya}
demonstrated that the room-temperature PL intensity of TFSI-passivated MoS$_2$ MLs grows by
around one to three orders of magnitude in comparison with non-treated (as-exfoliated) MoS$_2$ samples. 
This suggests that the quality of source MoS$_2$ crystals used for exfoliation, determined by their origin (in the case of natural crystals), the growth procedure, the number of defects $etc.$, strongly influences the resultant effect of the superacid treatment.
Moreover, it has been shown that the passivation of MoS$_2$ monolayers suppresses  
a dominant defect-related emission seen in their PL spectra measured at liquid helium temperature. 
However, the relative emission intensity due to the neutral and charged excitons seen at low temperature
is not affected by the treatment process~\cite{cadizacid}. In consequence, it can be expected 
that the results of the passivation process, $e.g.$ the increase of the PL intensity, may strongly
depend on initial sample's quality and/or experimental conditions. 

In order to deepen the understanding of chemical processing effects on S-TMD monolayers, we have carried out a study of optical properties of ML MoS$_2$ subjected to the superacid
treatment. We have not found any substantial increase in the room-temperature PL intensity caused by the passivation process. On the other hand, we have observed that the PL emission due to
the negative trion quenches because of the treatment. A systematic energy redshift of the neutral
excitons associated with both the A and B fundamental excitonic resonances at the K points of 
the ML MoS$_2$ Brillouin zone (BZ) is also apparent in our results.  We associate our observations with the passivation of 
unintentional doping centers at the surface of MoS$_2$ monolayer and the resulting decrease in the out-of-plane
electric field's magnitude in the sample. The change of the excitonic energy affects also the 
Raman scattering (RS) efficiency in the studied MLs. A substantial enhancement of the RS intensity is observed
when the excitation energy is resonant with the neutral exciton in the vicinity of the A resonance. 
On the contrary, the RS of light being in resonance with the corresponding trion can hardly be recognized in the recorded spectra.

\section{Samples and experimental setups \label{experiment}}

MoS$_2$ monolayers were  prepared  by mechanical exfoliation 
of bulk crystals (2H phase) using a two-stage Scotch-tape- and polydimethylsiloxane-based technique~\cite{gomez}.
After a non-deterministic transfer onto Si/(300 nm)SiO$_2$ substrates, the flakes of interest were first identified by 
visual inspection under an optical microscope and then cross-checked 
with Raman scattering and PL measurements at room temperature in order
to unambiguously determine their thicknesses. The samples 
containing MoS$_2$ MLs were then chemically treated in 
bis(trifluoromethane) sulfonimide (TFSI) following the procedure described in Ref.~\citenum{cadizacid}.
Several succesive treatment rounds were applied to the same sample.

The PL and Raman scattering measurements were carried out using $\lambda$=514.5 nm 
(2.41~eV) and $\lambda$=632.8 nm (1.96~eV) lines of a continuous-wave Ar-ion 
and He-Ne laser, respectively. The studied samples were placed on a cold 
finger in a continuous flow cryostat mounted on $x–y$ motorized positioners. 
The excitation light was focused by means of a 50x long-working distance 
objective with a 0.5 numerical aperture giving a spot of about 1 $\mu$m
diameter. The signal was collected via the same microscope objective, sent 
through a 0.5 m monochromator, and then detected using a liquid-nitrogen-cooled charge-coupled-device camera. The excitation power focused on the 
sample was kept at 50 $\mu$W during all measurements to avoid local heating. 
For RC study, the only difference in the experimental setup with respect to 
the one used for recording the PL and Raman scattering signals concerned the 
excitation source, which was replaced by a 100 W tungsten halogen lamp. The 
light from the lamp was coupled to a multimode fiber of a 50 $\mu$m core diameter, 
and then collimated and focused on the sample to a spot of about 4 $\mu$m diameter. 
We define the RC spectrum as $RC(E)=\frac{R(E)-R_0(E)}{R(E)+R_0(E)}\times 100\%$, 
where $R(E)$ and $R_0(E)$ are, respectively, the reflectances of the dielectric stack composed of an MoS$_2$ monolayer supported by an Si/SiO$_2$ substrate and of the Si/SiO$_2$ substrate alone.

\section{Experimental results \label{results}}

\subsection{Carrier concentration in a superacid treated monolayer \label{results:carriers}}

\begin{center}
	\begin{figure}[t]
		\centering
		\includegraphics[width=1\linewidth]{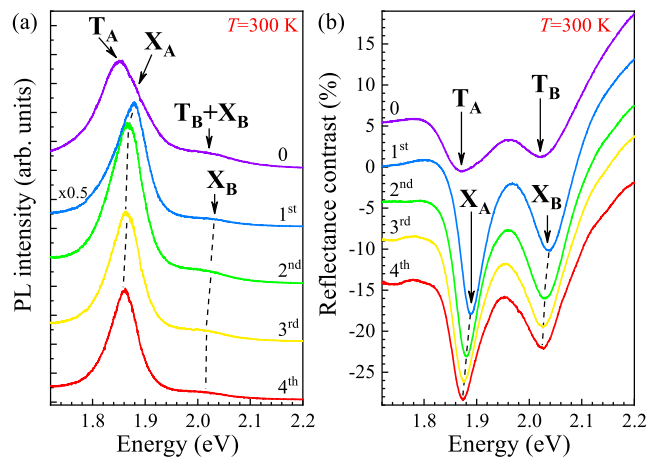}%
		\caption{Effect of successive TFSI passivations on (a) photoluminescence and (b) reflectance contrast spectra measured on an MoS$_2$ monolayer at room temperature. Denoted by '0' are the spectra recorded on an as-exfoliated ML, while labels '1$^\textrm{st}\dots$ 4$^\textrm{th}$' correspond to the number of passivations the ML was subjected to.}
		\label{fig:fig_1}
	\end{figure}
\end{center}

\begin{center}
	\begin{figure*}[t]
		\centering
		\includegraphics[width=1\linewidth]{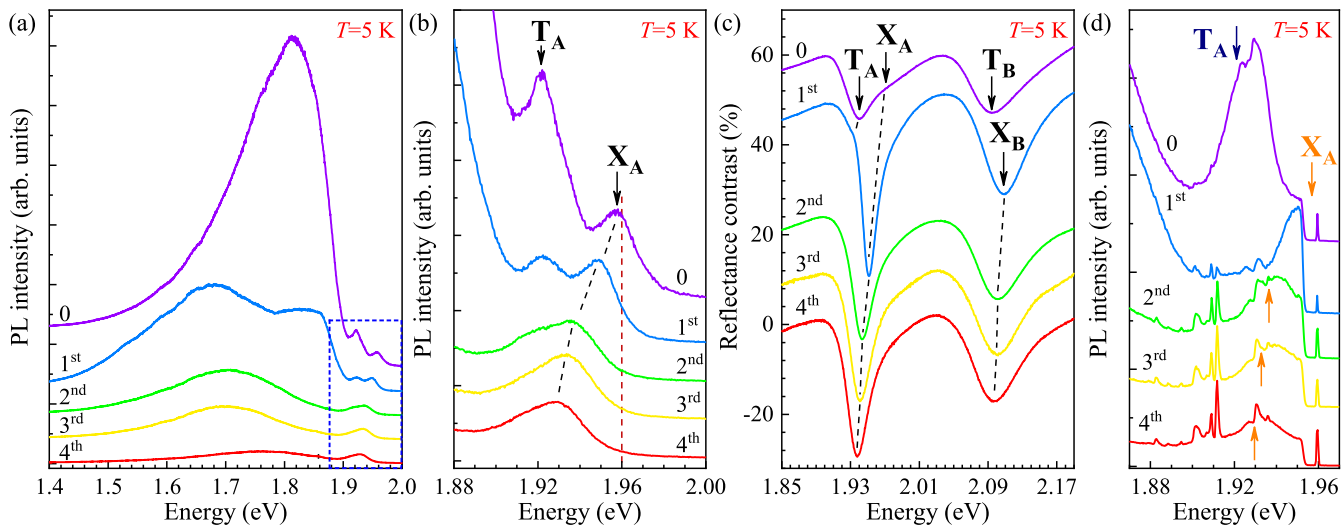}%
		\caption{Effect of successive TFSI passivations on low-temperature, $T$=5 K, (a),(b) photoluminescence, (c) reflectance contrast and (d) Raman scattering spectra measured on an MoS$_2$ monolayer at $\lambda$=632.8 nm excitation. Denoted by '0' are the spectra recorded on an as-exfoliated ML, while labels '1$^\textrm{st}\dots$ 4$^\textrm{th}$' correspond to the number of passivations the ML was subjected to. The red dashed vertical line in panel (b) indicates the energy of resonant Raman excitation conditions ($\lambda$=632.8 nm). The energies of T$_\textrm{A}$ and X$_\textrm{A}$ lines, inferred from the analysis of spectra shown in panel (b), are denoted by blue and orange arrows, respectively.}
		\label{fig:fig_2}
	\end{figure*}
\end{center}

Room-temperature ($T$=300~K) PL and RC spectra of an exfoliated MoS$_2$ monolayer before and 
after four successive TFSI passivations are presented in Fig.~\ref{fig:fig_1}. A comparison 
between the PL spectra measured on the as-exfoliated ML and after the first round of superacid treatment 
allows us to identify up to three emission lines denoted by T$_\textrm{A}$, X$_\textrm{A}$ 
and T$_\textrm{B}$+X$_\textrm{B}$ (see Fig.~\ref{fig:fig_1}(a)). We ascribe these peaks 
to the neutral (X$_\textrm{A}$, X$_\textrm{B}$) and charged (T$_\textrm{A}$ and T$_\textrm{B}$) 
excitons formed in the vicinity of the so-called A and B excitonic transitions near the fundamental band gap 
at the K$^\pm$ points of the BZ, which occur due to sizeable spin-orbit splitting in the valence band and smaller, though non-negligible, in the conduction band~\cite{splendiani2010}. As can be appreciated in Fig.~\ref{fig:fig_1}(a), 
the effect of superacid treatment on the PL spectrum is the most prominent after the first round. 
We observe a two-fold increase of the PL intensity as a consequence of the passivation, which stays
in contrast to previous reports~\cite{amaniScience,AmaniNano2016,AmaniACS2016,Kim2017,cadizacid} demonstrating at room temperature an improvement of PL intenisty by at least one order of magnitude. 
The second significant result of the first superacid treatment is a blueshift of the emission line related 
to the A exciton by about 30~meV. The corresponding RC spectrum of the as-exfoliated ML displays 
two resonance we denote by T$_\textrm{A}$ and T$_\textrm{B}$ in Fig.~\ref{fig:fig_1}(b). 
After the first passivation, a similar blueshift of about 20 meV is also visible in the RC spectra for both the A- (T$_\textrm{A}$, 
X$_\textrm{A}$) and B-exciton (T$_\textrm{B}$, X$_\textrm{B}$) resonances, and is accompanied by a gain in their intensities. Moreover, the 
blueshifts are followed by reduction of linewidths of the emission/absorption lines
(see Figs ~\ref{fig:fig_1}(a) and (b)). The observed changes in both the PL and RC spectra as a 
result of the first superacid treatment can be understood in terms of a significant decrease of
high non-intentional doping of the as-exfoliated monolayer. While mainly the charged excitons (T$_\textrm{A}$ 
and T$_\textrm{B}$) contribute to both the PL and RC features in the as-exfoliated sample, 
the neutral excitons (X$_\textrm{A}$ and X$_\textrm{B}$) dominate the PL and RC spectra after 
the first superacid treatment. In favour of our interpretation, the observed blueshifts of about 20-30~meV reasonably match the 
energy separation between the neutral and charged excitons (X$_\textrm{A}$ and T$_\textrm{A}$) in monolayer MoS$_2$, reported to date in several publications~\cite{mak2013,cadizacid,cadiz}. Moreover, it is well known that the oscillator strength of a neutral exciton is 
much bigger than that of a trion~\cite{arora2015,molasNanoscale}, which is reflected in the 
intensities of the corresponding resonances seen in the RC spectra plotted in Fig.~\ref{fig:fig_1}(b) and Fig.~\ref{fig:fig_2}(c). 
Note that an analogous effect is less visible for the B-exciton emission, probably due to its smaller intensity. The influence of successive passivations (starting from the second 
one) on the PL and RC spectra is less pronounced and mostly appears as monotonic redshifts of both the 
X$_\textrm{A}$ and X$_\textrm{B}$ resonances (the total shift amounts to about 10~meV after the fourth passivation round). This evolution can be explained in terms of modification of the built-in vertical electric 
field in the structure, $i.e.$ the quantum-confined Stark effect, due to passivation of defects 
on the sample surface. Note that the linewidth of the A-exciton emission, see Fig.~\ref{fig:fig_1}(a), 
is also reduced from $\sim$90~meV before the superacid treatment to $\sim$60~meV after the fourth passivation, which confirms the quenching of the trion emission and stays in agreement with similar results reported in Ref.~\citenum{cadizacid}.

The effect of superacid treatment described above can be studied in greater detail at low temperature
because linewidths of excitonic emission/absorption resonances in S-TMDs significantly decrease with reducing the temperature~\cite{arora2015,aroramose2,molasNanoscale}. Figure~\ref{fig:fig_2} presents a set of low-temperature ($T$=5~K) PL, RC and Raman scattering spectra measured on an MoS$_2$ ML before and after each of four successive passivations. As can be seen, both in the PL and RC spectra of the as-exfoliated ML, 
the X$_\textrm{A}$ and T$_\textrm{A}$ features are well 
resolved. This allows us to analyse the effect of superacid treatment more accurately. 
The PL spectrum of the as-exfoliated ML, shown in Fig.~\ref{fig:fig_2}(a),
is dominated by a broad emission band covering hundreds of meV's and commonly ascribed to defect states~\cite{cadizacid,molas}. 
An efficient quenching of this emission band with every succesive passivation step is clearly visible. Its 
maximum intensity is on a comparable level as the X$_\textrm{A}$ intensity after the fourth passivation. More information can be obtained by making a closer inspection of the low-temperature PL due to
the A-transition-related excitonic features (see Fig.~\ref{fig:fig_2}(b)). The observed peaks are attributed to the charged 
(T$_\textrm{A}$) and neutral (X$_\textrm{A}$) excitons. Similarly to the room-temperature behaviour (see Fig.~\ref{fig:fig_1}(a)), as a result of the superacid treatment the PL signal coming from the trion complex quenches, leaving 
the neutral-exciton emission as a main feature of the spectrum. 
A systematic redshift of the X$_\textrm{A}$ peak with successive treatment rounds can also be noticed 
in Fig.~\ref{fig:fig_2}(b). As we already proposed above, these two observations can be explained in terms of 
significant decrease of high non-intentional doping of the as-exfoliated monolayer, combined with a reduced influence of the quantum confined Stark effect reflecting the modification of the built-in vertical electric field in the structure 
due to passivation of defects on the sample surface. Both effects impact also the RC spectra 
in the energy range of the A and B excitons (see Fig.~\ref{fig:fig_2}(c)). In particular, the observed 
non-monotonic evolution of the A-exciton minimum in the RC spectrum supports the discussed scenario. 
A substantial density of charge carriers in the as-exfoliated monolayer results in a strong trion T$_\textrm{A}$ 
resonance which dominates the RC spectrum. Without any chemical treatment, the neutral exciton X$_\textrm{A}$ can only be recognized 
as a high-energy component of the trion minimum. After the first passivation, the contribution from the trion can still be observed as a low-energy shoulder of the X$_\textrm{A}$ dip, 
but further treatment rounds almost completely remove it from the spectrum. The reduction of the charge density 
and the evolution of the electric field in the sample also explain the behaviour of the RC minimum occuring in the B-exciton spectral range (see Fig.~\ref{fig:fig_2}(c)), which is analogous to that corresponding to the A exciton. However, due to larger linewidths of the B-exciton resonances, it cannot be seen that clearly. 
In particular, the jump of the B-exciton dip to a higher energy after the first treatment round should share the same origin with an almost identical shift observed for the T$_\textrm{A}$+X$_\textrm{A}$ feature. It means that the minimum in the RC spectrum corresponding to the B-resonance in 
the as-exfoliated ML is mostly composed of a contribution associated with the absorption of light by the charged exciton 
T$_\textrm{B}$. On the contrary, the RC features in the superacid-treated samples are due to the neutral excitons (X$_\textrm{B}$), 
which undergo a quantum confined Stark shift. It is worth to point out that the observed redshift resulting from the modification of the built-in vertical electric field in the ML is similar for both the X$_\textrm{A}$ and X$_\textrm{B}$ 
resonances and equals to 13~meV and 10~meV, respectively, as can be red from Fig.~\ref{fig:fig_2}(b). Note that an analogous 
effect to described above can also be seen in the PL experiments for the emission lines in the vicinity 
of the A-resonance, X$_\textrm{A}$ and T$_\textrm{A}$, presented in Fig.~\ref{fig:fig_2}(a) and (b). In this case, 
however, the emission originating from the trion  complex does not fully vanish from the PL spectra even after four passivations, 
probably due to a small residual doping which suffices to observe the trions in the photoluminescence, but not in absorption-like measurements.

One of pertinent questions regarding the results of optical experiments discussed above is the sign of the charged exciton resonances. In general, a trion, as a complex of an 
electron-hole pair and an extra carrier (electron or hole), can be negative (two electrons + a hole) or
positive (an electron + two holes). In our case, an identification of which of these situations we are dealing with cannot be done in an unambiguous way. Unfortunately, the same effects of the superacid treatment as observed in this work for the resonances in the 
vicinity of the A and B excitons might be expected for both negative and positive trions.

\subsection{Resonant Raman scattering - the effect of the charge state \label{results:raman}}

After four superacid treatment rounds, the X$_\textrm{A}$ energy cumulatively downshifts by about 30 meV, see Fig.~\ref{fig:fig_2}(d), becoming comparable to the T$_\textrm{A}$ energy in the as-exfoliated MoS$_2$ ML. This evolution creates a unique opportunity 
to study Raman scattering under variable resonance conditions using a fixed excitation wavelength, in our case the $\lambda$=632.8~nm line of a He-Ne laser, usually employed for this kind of investigations in MoS$_2$. 
It has previously been shown that excitation at this wavelength leads to resonant enhancement of the A$_{1\textrm{g}}$ line in bulk 
MoS$_2$~\cite{frey} and to multiphonon Raman scattering~\cite{golasaAPL}, both observed at low temperature. 
The effect of excitation energy on the optical response of ML MoS$_2$ was also studied~\cite{lee2015,placidi2015,carvalho2015,kutrowska2018}, 
but the concept of carrying out such measurements under different resonant conditions varied by changing the carrier density in the ML has not been widely explored. Our work attempts to fill this gap.

\begin{center}
	\begin{figure}[t!]
		\centering
		\includegraphics[width=1\linewidth]{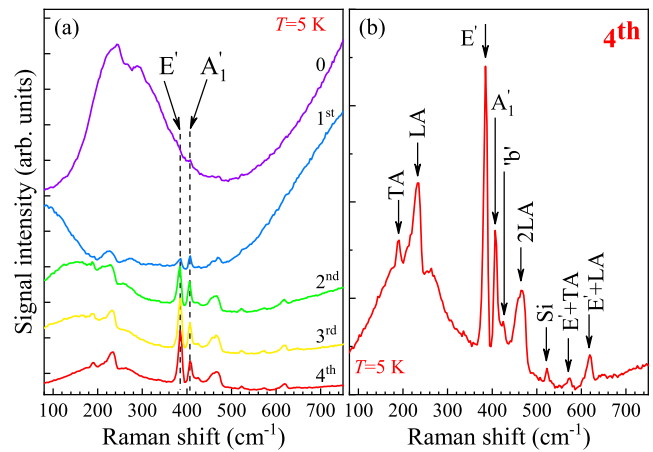}%
		\caption{(a) Effect of successive TFSI passivations on the Raman scattering spectrum measured under $\lambda$= 632.8 nm excitation on an MoS$_2$ monolayer at $T$=5 K. Denoted by "0" is the spectrum recorded on an as-exfoliated ML, while labels '1$^\textrm{st}\dots$4$^\textrm{th}$' correspond to the number of passivations the ML was subjected to. (b) Low-temperature ($T$=5 K) Raman scattering spectrum measured under $\lambda$= 632.8 nm excitation on an MoS$_2$ monolayer which underwent 4 superacid passivation rounds. Unless stated otherwise in the main text, the indicated phonons  correspond to the M point of the Brillouin zone.}
		\label{fig:fig_3}
	\end{figure}
\end{center}

The effect of superacid treatment on the low-temperature emission spectrum of ML MoS$_2$ excited with $\lambda$=632.8~nm light is shown in Fig.~\ref{fig:fig_2}(d). As indicated by orange arrows, such excitation energy matches 
the neutral exciton energy in the as-exfoliated sample. At lower energy an emission related to 
the T$_\textrm{A}$ trion can be recognized as a structure superimposed on a broad peak centered at $\sim$1.921~eV.
As discussed earlier in the present paper, the trion disappears from the spectrum as a result of
superacid treatment. This evolution of the spectrum shape is accompanied by a redshift of the neutral exciton. Moreover, with successive passivations, the structure due to Raman scattering becomes more and more visible. The spectra shown in Fig.~\ref{fig:fig_2}(d), are also presented in Fig.~\ref{fig:fig_3}(a) in a more conventional way, $i.e.$ as a function of the Raman shift. The most intense Raman peaks in ML MoS$_2$ originate from two Raman-active modes: E$^{'}$, which results from in-plane vibrations of the two S atoms with respect to the Mo atom, and A$^{'}_1$ which is due to out-of-plane and out-of-phase vibrations of the S atoms~\cite{Lee2010}. Surprisingly, the Raman scattering signal measured at $T$=5 K on the as-exfoliated monolayer is extremely weak, even though the excitation energy almost equals the X$_\textrm{A}$ 
energy (see Fig.~\ref{fig:fig_2}(c)). This leads to conditions under which the energies of the trion emission
and of the first order Raman scattering processes coincide, or, in other words, come into resonance. 
When the monolayer is subjected to successive superacid treatments, the intensity of the Raman scattering 
signal significantly increases (for the A$^{'}_1$ line, the enhancement amounts to about one order of magnitude). 
Another striking feature of the low-temperature spectra is the presence of an emission band at $\sim$200~cm$^{-1}$, see Fig.~\ref{fig:fig_3}(a). 
There are no center-zone Raman modes expected for ML MoS$_2$ in that energy range. The lineshape of the emission 
band corresponds, however, to the total integrated density of phonon states in the sample~\cite{molina}. 
This suggests that scattering involving single-phonon processes from outside the center of the BZ are allowed in this case. Two maxima, which are particularly well visible in the spectrum presented in Fig.~\ref{fig:fig_3}(a), 
correspond to the transverse acoustic (TA) and longitudinal acoustic (LA) phonons near the M point from 
the border of the BZ. Most likely explanation of their presence in the Raman scattering spectrum points at disorder 
in the sample which localizes phonons~\cite{golasaACTA,golasaAIP,mignuzzi}. Such interpretation does not support, however, the substantial enhancement of the Raman peaks 
seen at low temperature in the spectra recorded on MoS2$_2$ monolayers treated with superacid solution several times. The comparison between spectra measured on as-exfoliated sample with the trion- and defect-dominated emission and on the sample after four passivation rounds clearly suggests that the enhancement is associated with bringing the Raman features into resonance with the neutral exciton. 

\begin{center}
	\begin{figure}[t!]
		\centering
		\includegraphics[width=1\linewidth]{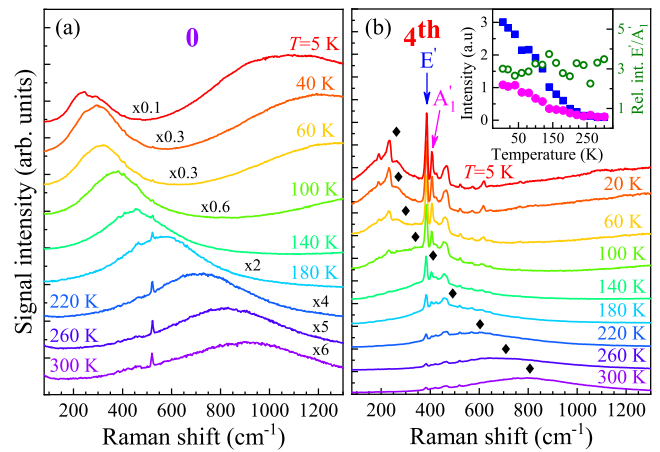}%
		\caption{Temperature evolution of resonant Raman scattering signal measured under $\lambda$= 632.8 nm excitation on an (a) as-exfoliated (0) and (b) four-times-passivated (4$^{\textrm{th}}$) MoS$_2$ monolayer. The spectra in panel (a) are multiplied by the factors specified above the experimental curves. The black diamonds in panel (b) indicate the X$_\textrm{A}$ energy inferred from the PL spectra measured under $\lambda$=514.5 nm excitation. The temperature evolution of the E$^{'}$ and A$^{'}_1$ modes intensities and their relative intensity are shown in the inset to panel (b).}
		\label{fig:fig_4}
	\end{figure}
\end{center}

Another possibility to tune the resonance conditions for the Raman scattering is by varying sample's temperature. The temperature evolution of optical spectra measured on the as-exfoliated ML and after four passivations 
is presented in Fig.~\ref{fig:fig_4}. As can be noticed in Fig.~\ref{fig:fig_4}(b), the intensities of all Raman scattering peaks measured on the sample subjected to four treatment rounds gradually decrease with increasing temperature.
We ascribe this effect to the shift of the neutral exciton energy away from the energy of the laser light due to
the temperature dependence of the band gap~\cite{arora2015,aroramose2,molasNanoscale}. Surprisingly, the 
quenching affects both the A$^{'}_1$ and E$^{'}$ modes. This is in contrast to the reported evolution of 
the Raman scattering efficiency in MoS$_2$ ML in which distinct exciton-phonon coupling strengths for the two modes can be explained by considering their symmetries with respect to the symmetries of the orbitals associated with 
the A, B and C (a high energy transition at around 2.7 eV) excitons~\cite{carvalho2015}. Our results 
show that at low temperature, the in-plane E$^{'}$ mode in ML MoS$_2$ undergoes a resonant enhancement
despite the symmetry effect on the exciton-phonon interaction. Additional information can be gained from 
the analysis of the temperature-driven shift of the neutral exciton, X$_\textrm{A}$,  energy. Its energy, 
as taken from the emission spectra excited at 514.5 nm, is denoted with diamonds in Fig.~\ref{fig:fig_4}(b). 
No significant enhancement of the Raman-scattering-related emission spectra occurs for the outgoing 
resonance between the excitonic energy and the energy of scattered light. A monotonic 
decrease of the emission intensity is rather related to detuning the excitation laser energy off the neutral exciton energy. This may suggest a possible scenario responsible for the observed effects. At low temperature, the laser energy in use leads to the formation of neutral excitons at higher k-vectors, which then relax to the minimum energy state at the K$^\pm$ points by emitting discrete phonon modes ($e.g.$ A$^{'}_1$). With increasing temperature, the excitonic energy shifts away from the laser energy 
and the relaxation involves mainly acoustic phonons which do not contribute to the discrete emission spectrum observed at low temperature. As can be seen in Fig.~\ref{fig:fig_4}(a), the process is not effective in a sample with high carrier density. In such a case, mostly acoustic phonons assist the relaxation processes 
and the discrete structure of the optical emission spectra can hardly be seen. Our observations point out 
the difference between the electron-phonon interactions with the neutral and charged excitons, as previously 
reported for monolayer WS$_2$~\cite{molasSR}. Furthermore, we showed that the carrier concentration in the studied MoS$_2$ monolayer plays a significant role for the intensity of the resonant Raman scattering signal. It is clear that our qualitative explanation should be supported by more strict theoretical analysis which is beyond the scope of this experimental work. We do believe, however, that our results can contribute to the understanding of fundamental physical processes in S-TMDs, which is of prime interest for their potential applications.

\section{Conclusions}
In conclusion, we have studied the effect of superacid (TFSI) treatment on the optical properties of monolayer MoS$_2$
with the aid of photoluminescence, reflectance contrast and Raman scattering spectroscopy employed in a broad temperature range. We have observed that the defect-related low-energy photoluminescence progressively disappears from the spectrum 
as a result of successive TFSI passivations. Moreover, we have found that the treatment results in systematic quenching of the charged exciton emission/absorption and the redshift of the neutral exciton emission/absorption associated with both the A and B resonance transitions at the K$^\pm$ points of the monolayer MoS$_2$ BZ. Furthermore, our results demonstrate that the charge state of dominant excitons in ML MoS$_2$ affects the Raman scattering under resonant excitation conditions. In particular, it has been shown that much more effective scattering occurs for the resonance with the neutral exciton than with the trion.

\section*{Acknowledgements}
The work has been supported by the European Research Council (MOMB project no. 320590), the EC Graphene Flagship project (no. 604391), the National Science Center (grants no. DEC-2013/11/N/ST3/04067, DEC-2015/16/T/ST3/00496, UMO-2017/24/C/ST3/00119, UMO-2017/27/B/ST3/00205), the Nanofab facility of the Institut N\'eel, CNRS UGA, and the ATOMOPTO project (TEAM programme of the Foundation for Polish Science co-financed by the EU within the ERDFund).

\bibliographystyle{apsrev4-1}
\bibliography{biblio_MoS2_Acid}

\end{document}